\documentclass[Physsubmission, Phys]{SciPost}

\hypersetup{
    colorlinks,
    linkcolor={red!50!black},
    citecolor={blue!50!black},
    urlcolor={blue!80!black}
}

\usepackage[bitstream-charter]{mathdesign}
\DeclareSymbolFont{usualmathcal}{OMS}{cmsy}{m}{n}
\DeclareSymbolFontAlphabet{\mathcal}{usualmathcal}

\begin{document}

\begin{center}{\Large \textbf{
Using coherent dipion photoproduction to image gold nuclei\\
}}\end{center}

\begin{center}
Spencer R. Klein\textsuperscript{1} for the STAR Collaboration

\end{center}

\begin{center}
{\bf 1} Nuclear Science Division, Lawrence Berkeley National Laboratory, Berkeley CA USA\\
* srklein@lbl.gov
\end{center}

\begin{center}
\today
\end{center}


\definecolor{palegray}{gray}{0.95}
\begin{center}
\colorbox{palegray}{
  \begin{tabular}{rr}
  \begin{minipage}{0.1\textwidth}
    \includegraphics[width=22mm]{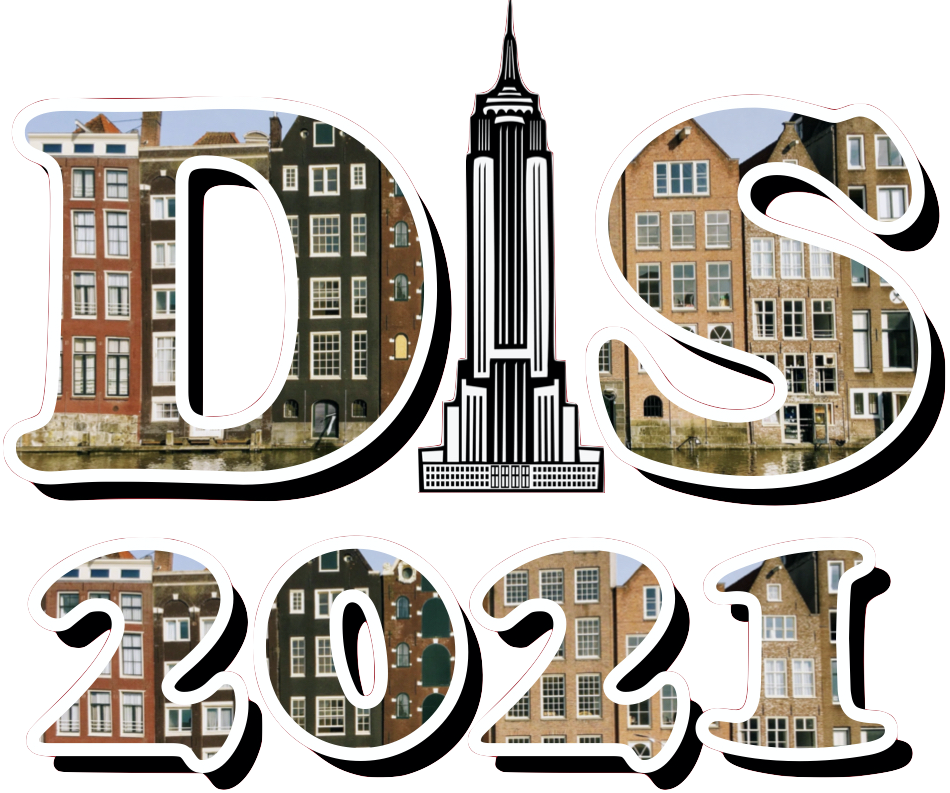}
  \end{minipage}
  &
  \begin{minipage}{0.75\textwidth}
    \begin{center}
    {\it Proceedings for the XXVIII International Workshop\\ on Deep-Inelastic Scattering and
Related Subjects,}\\
    {\it Stony Brook University, New York, USA, 12-16 April 2021} \\
    \doi{10.21468/SciPostPhysProc.?}\\
    \end{center}
  \end{minipage}
\end{tabular}
}
\end{center}

\section*{Abstract}
{\bf

Vector meson photoproduction offers the opportunity to image target nuclei.  The two-dimensional Fourier transform $d\sigma_{\rm coherent}/dt$ of coherent vector meson photoproduction gives the two-dimensional distribution of interaction sites in the target.  Since vector meson photoproduction occurs, at lowest order, via two-gluon exchange, this is sensitive to gluon shadowing.  We present an analysis of  $\pi^+\pi^-$ photoproduction using data from the STAR detector and a study of $d\sigma_{\rm coherent}/dt$, with an emphasis on probing the nuclear shape and its systematic uncertainties.
}


\section{Introduction}
\label{sec:intro}

Vector meson photoproduction has long been used as a probe of nuclei \cite{Alvensleben:1970uw}.   The photon fluctuates to a quark-antiquark dipole which scatters hadronically (but elastically) with the target.   In lowest order perturbative QCD (pQCD),  the elastic scattering proceeds via the exchange of two gluons, so it is a useful probe of the gluon content of nuclear targets.   High-energy photoproduction on proton targets was extensively studied at HERA.   Unfortunately, HERA did not accelerate $A>1$ nuclei, so high-energy photoproduction studies on nuclear targets had to await the advent of ultra-peripheral collisions at RHIC and the LHC.   There, studies of $\rho$ photoproduction on gold and lead targets pointed to the importance of high-mass intermediate states {\it i. e.} the Glauber-Gribov formalism was required to properly describe $\rho$ photoproduction; a straight Glauber calculation overpredicts the data \cite{Frankfurt:2015cwa}.   Data on $J/\psi$ production on lead targets at the LHC supports the presence of moderate shadowing, beyond what is predicted by a Glauber calculation \cite{Acharya:2021ugn}.

Photoproduction can go beyond simple measurements of gluon abundance, though.  In the Good-Walker paradigm \cite{Good:1960ba,Klein:2019qfb}, $d\sigma_{\rm Coherent}/dt$ is related to the transverse distribution of interaction sites (the average nuclear configuration), while $d\sigma_{\rm incoherent}/dt$ is related to instantaneous (event-by-event) fluctuations in the nuclear configuration, including the positions of the nucleons and partonic fluctuations, such as gluonic hot spots.

Measurement of the transverse nuclear profile in UPCs can be problematic, because the measured transverse momentum ($p_T$) spectrum includes components from the photon $p_T$ and due to the detector resolution, as well as the nuclear scattering.   Here, we explore a different approach, seeing how well $d\sigma_{\rm Coherent}/dt$ can be fit by a model that includes scattering from a target nucleus that is treated as a linear combination of a Woods-Saxon nucleus (no saturation effects whatsoever) and a black disk (fully saturated). 

\section{The STAR detector and the dataset}

\begin{figure}[h]
\centering
\includegraphics[width=0.5\textwidth]{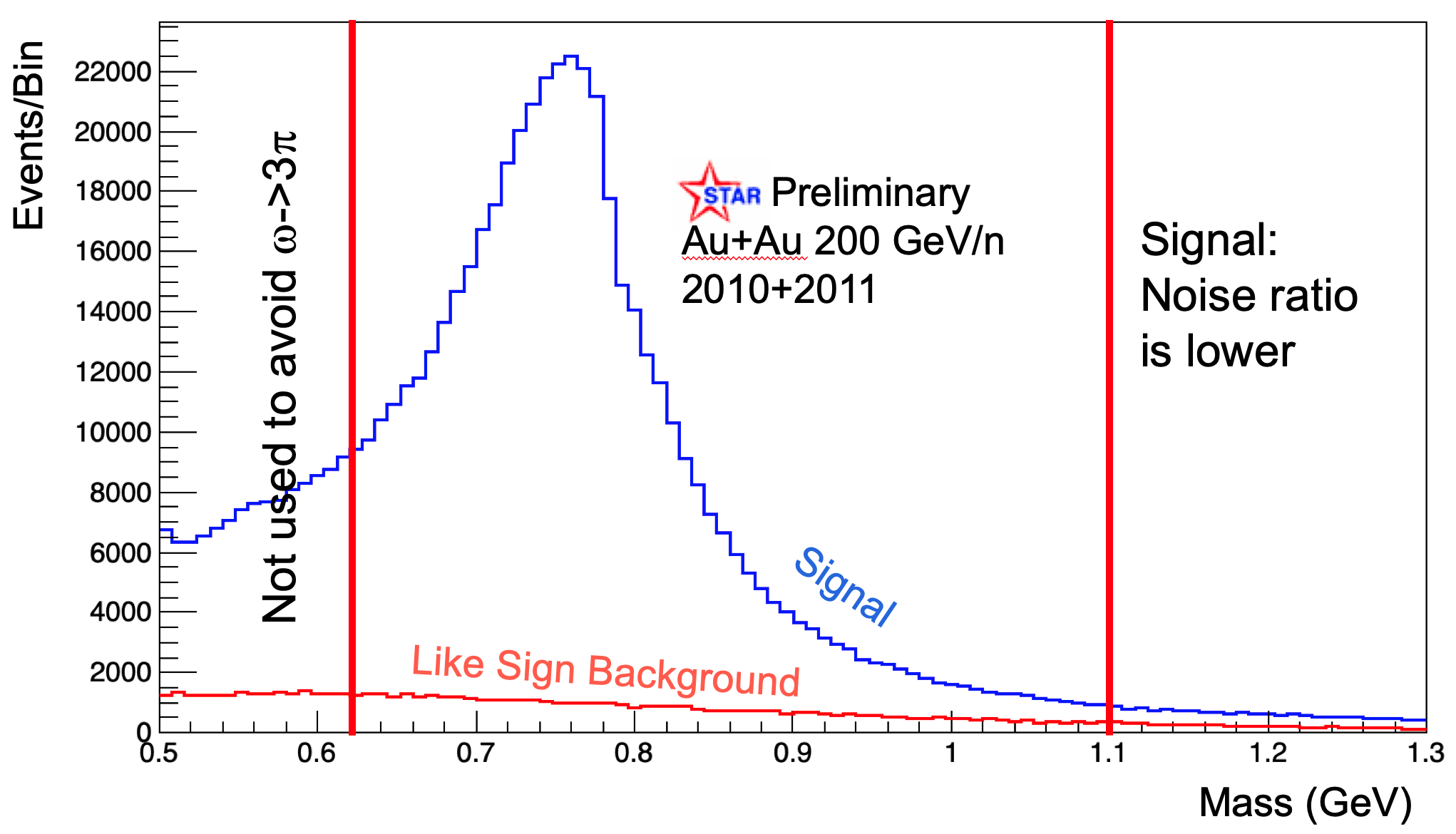}
\caption{Mass spectrum for unlike-sign and like-sign dipion pairs.}
\label{fig:mass}
\end{figure}

This analysis uses data collected with the STAR detector during the 2010 and 2011 running.  For this analysis, the main detector elements were a cylindrical time projection chamber (TPC) and a time-of-flight (TOF) system in a 0.5 T solenoidal magnetic field, and two zero degree calorimeters (ZDCs) which detected neutrons from nuclear breakup.  The trigger required 2-6 hits in the time-of-flight system, plus neutron signals in both ZDCs, while the analysis required exactly two tracks with at least 25 hits in the TPC.  The vertex was required to be within 50 cm in $z$ of the center of the TPC, and the pion pair was required to have pair $|{\rm rapidity}|>0.04$, to remove cosmic-ray muons which might mimic a pair.  Pairs were required to have an invariant mass greater than 0.62 GeV, to remove background from photoproduced $\omega\rightarrow\pi^+\pi^-\pi^0$.   The maximum mass was chosen to be 1.1 GeV.  At higher masses, the signals are smaller, and the signal:background ratio falls.  There are 635,917 unlike-sign pairs and 71,187 like-sign pairs in the full histogram, giving a signal:background ratio of about 9:1.  Figure \ref{fig:mass} shows the mass spectrum for unlike- and like- sign pairs.   The mass spectrum is well fit by a combination of $\rho\rightarrow\pi^+\pi^-$, direct $\pi^+\pi^-$ and $\omega\rightarrow\pi^+\pi^-$, with ratios that are very similar to earlier STAR work \cite{Adamczyk:2017vfu}.  

Although it may seem strange to require nuclear breakup while studying coherent photoproduction, most neutron emission comes from nuclear excitation caused by the exchange of additional photons (beyond the photon that produced a dipion).   These additional photons are independent of the dipion production, except for their common impact parameter.   Earlier STAR studies demonstrated that the additional photons do not interfere with coherent production \cite{Abelev:2007nb}, although they do bias the reaction toward smaller $\langle b\rangle$ \cite{Baltz:2002pp,Baur:2003ar}.

\begin{figure}[h]
\centering
\includegraphics[width=0.48\textwidth]{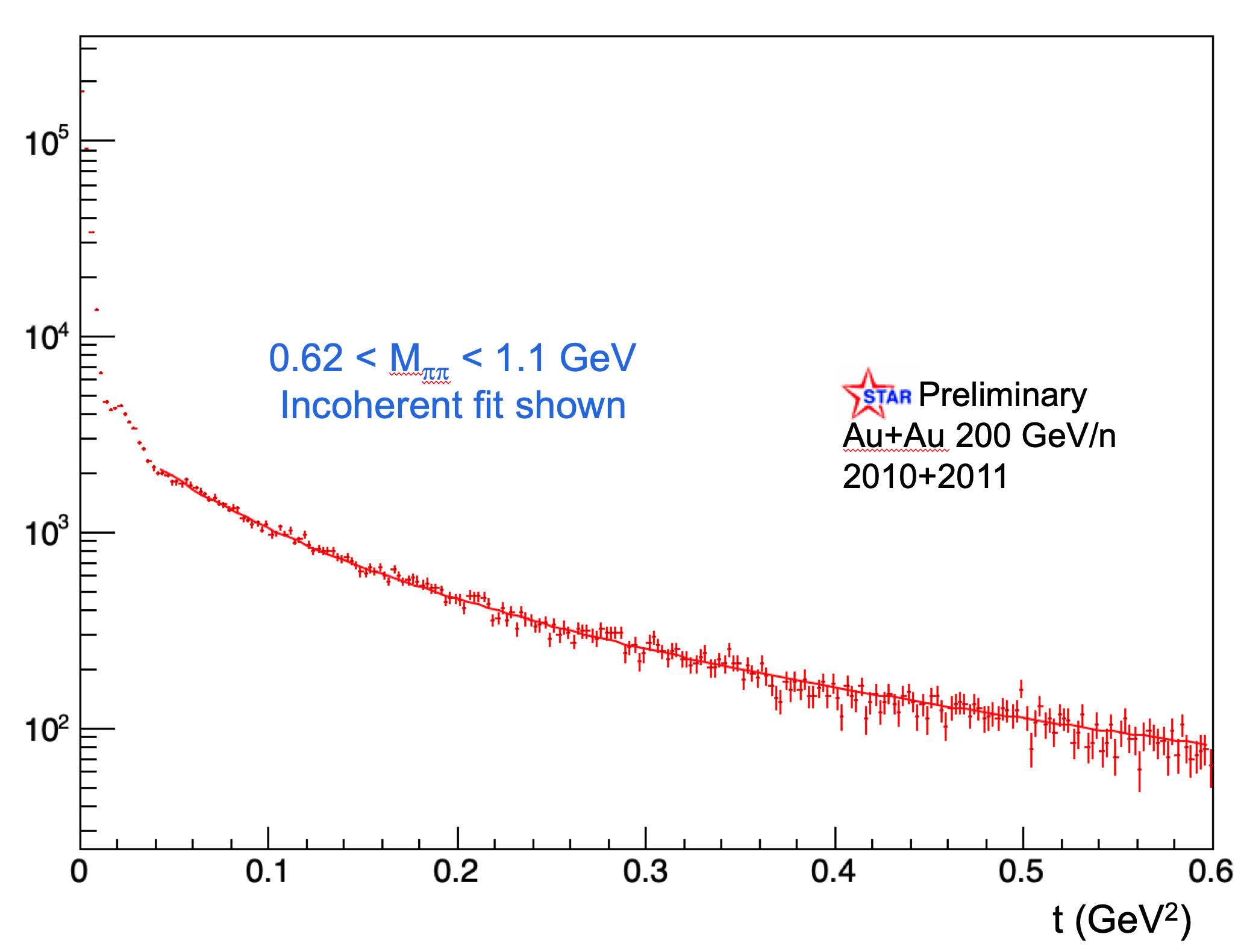}
\includegraphics[width=0.49\textwidth]{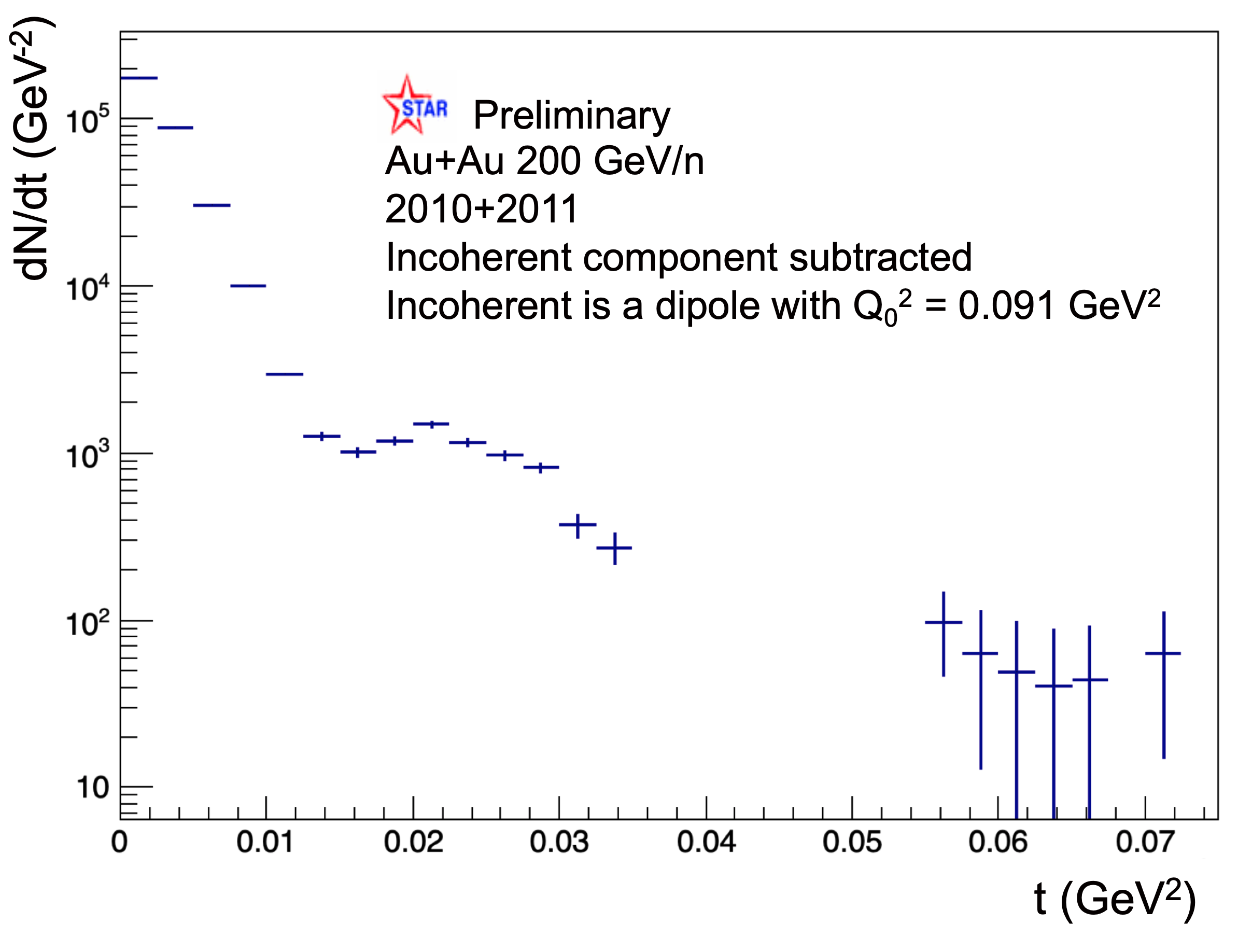}
\caption{(a) $d\sigma/dt$ for dipion pairs with 0.62 $< M_{\pi\pi} < 1.3$ GeV, with the dipole fit shown by the solid red line. (b) $d\sigma/dt$, after subtraction of the coherent contribution, with an expanded $t$ scale, showing the coherent result.}
\label{fig:dsdt}
\end{figure}

The first analysis step is to subtract the incoherent contribution to $d\sigma/dt$ ($t$ is the usual Mandelstaam $t$), leaving the coherent contribution.   We find the incoherent contribution by fitting $d\sigma/dt$ at large $|t|$ where the coherent contribution is small, $0.05 < |t| < 0.45$\ GeV$^2$.
The incoherent contribution is fit with a dipole form factor
\begin{equation}
\frac{d\sigma}{dt} = \frac{A/Q_0^2}{(1+|t|/Q_0^2)^2}.
\end{equation}
The fit finds $Q_0=302.5\pm2.5$ MeV, with a $\chi^2/DOF$ of $160/158$, similar to the $Q_0=314^{+0.023}_{-0.025}$~MeV found in the previous STAR work \cite{Adamczyk:2017vfu}.   This is consistent with the expectations for recoil from a single proton.  Figure \ref{fig:dsdt} shows $d\sigma/dt$ along with the fit.  An exponential function, used in some earlier analyses, would not be a good fit to the data.  With the log scale on the $y$ axis of Fig. \ref{fig:dsdt}, an exponential function would appear as a straight line.

This subtraction lead to  $d\sigma_{\rm coherent}/dt$, as shown on the right panel of Fig. \ref{fig:dsdt}.  Around the second minimum, $t \approx 0.05$ GeV$^2$, the subtraction returns negative values (not shown on the plot).  This may indicate that the dipole formula fails for smaller $t$, possibly due to the small energy transfer to the nucleus. This fit is compatible with, but slightly below the fit in the 2017 STAR paper \cite{Adamczyk:2017vfu}, due to the slightly different $t$ range used here.

If incoherent photoproduction occurs when a Pomeron recoils against a single nucleon (as suggested by the dipole fit), then the energy transfer is related to the momentum transfer $E=t/2m_p$.  The minimum energy to eject a neutron or a proton from a gold nucleus is  8.07 MeV or 5.27 MeV, corresponding to momentum transfers of 122 MeV/c and 99 MeV/c, or $t\approx 0.01$ This is below the second minimum, but some threshold behavior is expected, and either the single-nucleon-recoil paradigm must fail, or the nucleon emission channels must drop out for $t<0.01$ GeV$^2$.   Photon emission via nuclear deexcitation is allowed at lower $t$, but is expected to account for only a small fraction of the total incoherent cross-section.
 
 \subsection{Shape Fits and Templates}
 
Previously, STAR made a two-dimensional Fourier transform of $d\sigma/dt$ to determine $F(b)$, the transverse profile of the interaction sites within the target - the heavy-ion equivalent of a generalized parton distribution for gluons.  However, that transform can introduce significant uncertainties.  Fourier transforms are exact for the full range $0 < p_T < \infty$, but the data has a limited $p_T$ range.  Imposing a maximum $p_T$ range introduces windowing artifacts \cite{Klein:2018grn}.  The measured $d\sigma/dt$  includes contributions from the Pomeron $p_T$, photon $p_T$, and the detector resolution.  The latter two components need to be removed to accurately probe the gluons.    They can be removed by unfolding  \cite{Acharya:2021bnz}, but this requires an accurate knowledge of both components, and can increase the uncertainties. 
 
 Here, we present an alternate approach, generating $p_T$ templates that include all three components.   We will do this for two different nuclear models - a Woods-Saxon nucleus, representing our expectations for a small dipole with a small interaction probability, and the other limit, which treats the nucleus as a black disk.  We will then fit the data to a linear combination of these two templates, as a measure of saturation in the target; higher saturation should correspond to a more black-disk-like nucleus.
 
 We treat the three components as uncorrelated, and add the $\vec{p_T}$ with a random azimuthal angle.  The components are normalized to have an integral of 1.  The resolution in $p_T$ can be represented with a Gaussian distribution, with $\sigma = 6$ MeV/c \cite{Abelev:2007nb}.  The photon $p_T$ distribution is given by \cite{Vidovic:1992ik,Klein:1999gv}
 \begin{equation}
 \frac{dN}{dp_T} \propto \frac{F^2(p^2) p_T^2}{p^2},
 \label{eq:photonpt}
 \end{equation}
 where $F(p^2)$ is the nuclear form factor, $p^2=p_T^2+p_z^2/\gamma^2$,  $p_z$ is the longitudinal momentum transfer to the nucleus and $\gamma$ is the nuclear Lorentz boost.  The $p_z$ term has a two-fold ambiguity regarding photon energy vs. rapidity.  Fortunately, it is small, and we can neglect it here.
 
 Equation \ref{eq:photonpt} is exact only if the photon spectrum is integrated from impact parameter $b=0$ to infinity.  The requirement that there be no hadronic interactions limits this data to roughly $b>2R_A$ while the requirement of mutual Coulomb dissociation biases it toward smaller impact parameters \cite{Baur:2003ar}.   Although it is possible to relate $\langle p_T^2 \rangle$ to $b$, there is no model-independent way to determine the photon $p_T$ distribution for limited impact parameter ranges \cite{Klein:2020jom}.   So, we will treat this as a poorly-known systematic error. 
 
 For the Woods-Saxon nuclear distribution, we use the analytic form of a hard-sphere nucleus convoluted with a Yukawa potential, with $p=p_T$ \cite{Klein:1999gv}
 \begin{equation}
 \frac{dN}{dp} \propto F^2(p^2) \propto \big(\big[\sin(pR_A)-pR_A \cos(pR_A)\big]\big[\frac{1}{1+a^2p^2}\big]\big),
 \label{eq:ws}
  \end{equation}
 where $R_A$ is the nuclear radius and $a=0.7$ fm is the range of the Yukawa potential.

 We also use Eq. \ref{eq:ws} as the form factor for the photon $p_T$, Eq. \ref{eq:photonpt}.  There, we take $R_A=6.38$ fm; this is the radius of the protons in the gold nucleus.  For the Pomeron form factor, we use $R_A=6.63$ fm, with the extra 0.25 fm accounting for the likely neutron skin of gold nuclei.   This Woods-Saxon approach ignores longitudinal coherence, and corresponds to something close to the impulse approximation, rather than a Glauber calculation.  
  
 The black-disk nuclear distribution is also represented analytically:
 \begin{equation}
 F(p) \propto \frac{2J_1(pR_A)}{pR_A}.
 \label{eq:bd}
 \end{equation}
 For the black disk, there is no unique $R_A$; the choice of the edge of the nucleus corresponding to an assumed rapid drop to zero density is somewhat arbitrary.  Here, we will choose $R_A=8$ fm.  This is a rather large value, but, as we will see, the fit prefers a large radius.   Equations \ref{eq:ws} and \ref{eq:bd} have one significantly difference between them; in Eq. \ref{eq:ws}, the zeros are linearly spaces, while in Eq. \ref{eq:bd}, they are not.  So, even if one lined up the first minimum by choosing appropriate nuclear radii, the higher minima would fall in different places, and a linear combination of the two functions would have too many minima. 
 
 Figure \ref{fig:fitresults} (left) shows the different components used in the templates: detector resolution, photon $p_T$, and the Woods-Saxon and black-disk models.  The resolution is relatively unimportant, dropping off at even moderate $p_T$.  The photon $p_T$ has more effect than the resolution, but still drops off substantially faster than either nuclear form factors.  It is enough, however, to largely fill in the diffractive minima.  At large $p_T$, the black disk form factor is significantly above the Woods-Saxon model.  Essentially, the black disk has a hard edge, which leads to larger harmonics.  So, $d\sigma/dt$ at large $|t|$ should be sensitive to the nuclear density profile, especially at the edges of the nucleus.   
 
 \section{Fitting and results}
 
\begin{figure}[t]
\centering
\includegraphics[width=0.49\textwidth]{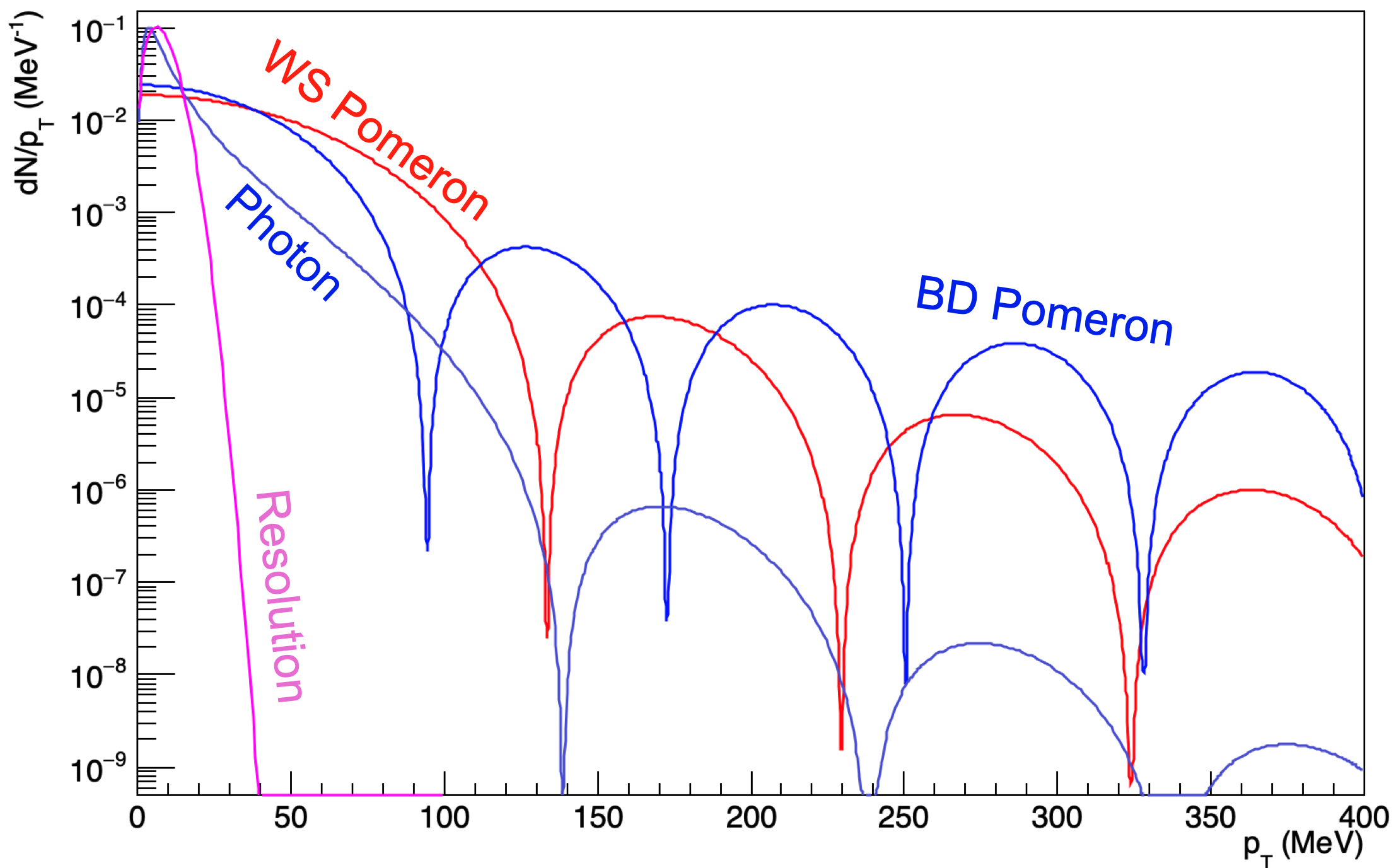}
\includegraphics[width=0.49\textwidth]{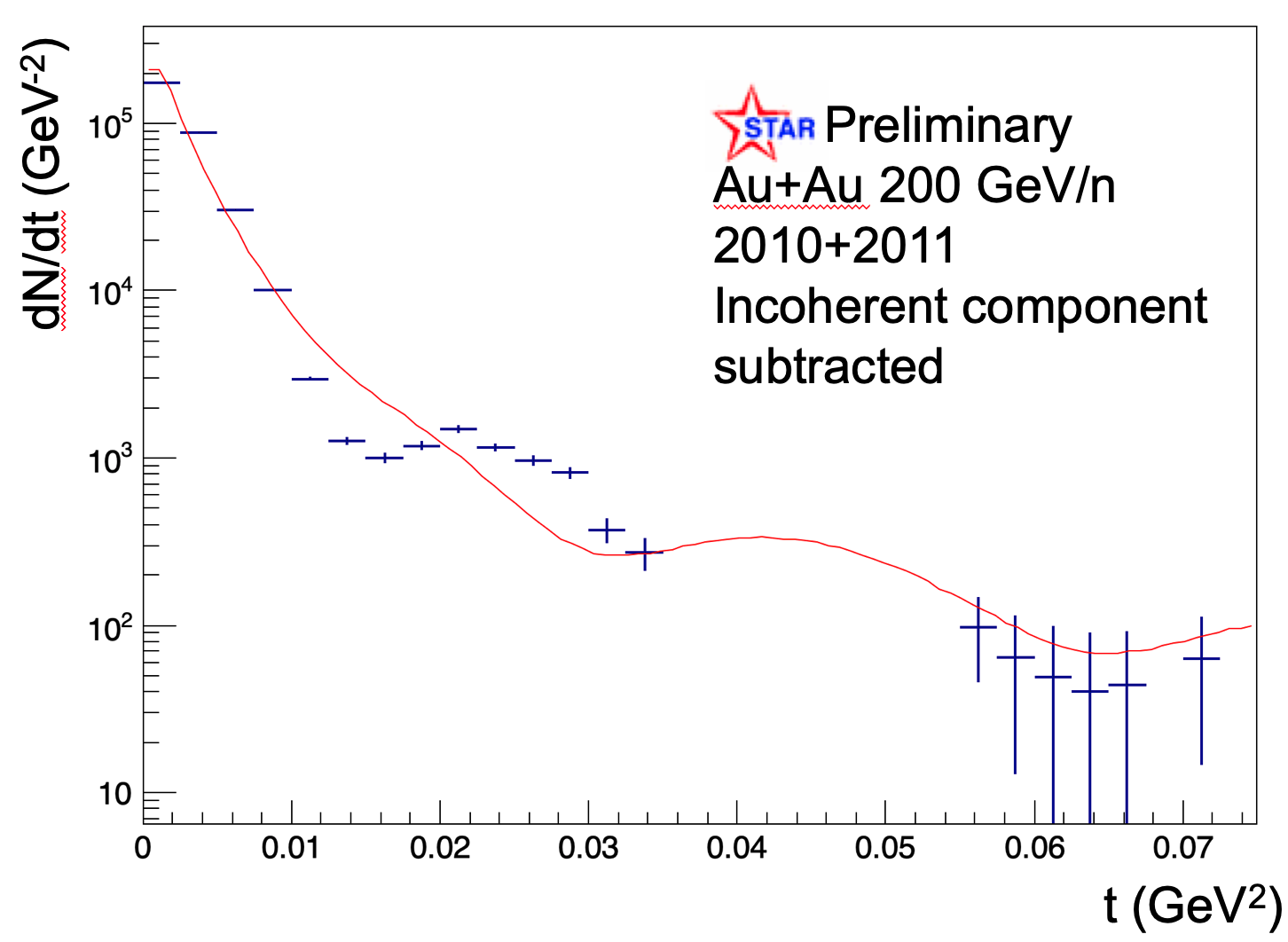}
\caption{(left) The components of the fitting template, for the detector resolution, photon $p_T$, and the Woods-Saxon and black-disk models. (right) The measured $d\sigma_{\rm coherent}/dt$, with the fit results.}
\label{fig:fitresults}
\end{figure}

 Figure \ref{fig:fitresults} (right) shows the fit results.  The best-fit value consists of  $\lambda=0.71\pm 0.01$ Woods-Saxon, with the remainder black disk.  However, the $\chi^2/$DOF\ $=224770/28$ - a terrible fit, showing that the model does not match the data.  The problem is that the fit would prefer an unphysically large nuclear radius of 9.5 to 10 fm. One factor that could possibly contribute to the nuclear radius would be the presence of Coulomb breakup.   If the breakup occurred before the photoproduction, it could increase the nuclear radius.  However, breakup is a lower-energy process, so should occur on longer time scales.   This radius mismatch dominates the fit, so the returned $\lambda$ is not trustworthy.  The radius is mostly determined by the slope of $d\sigma/dt$ below the first minimum, where most of the events are.  This radius-mismatch also pushes the first diffractive minimum in the fit out to much higher $t$ than in the data; a larger radius would move the dip to the left.  
 
An alternative approach, inspired by the dipole model, would be to fit to the square of the integrated (along $z$) density profile; the square being to account for two-gluon couplings to the target.   However, at the relevant $Q^2$ ($Q^2\approx M_{\pi\pi}^2$), it is unclear if a model that is sensitive to the partonic constituents of the target is appropriate.  
 
\section{Conclusion}

We have attempted to fit $d\sigma_{\rm coherent}/dt$ for $\pi^+\pi^-$ photoproduction to linear combination of that expected for weakly interacting (small) dipoles and for strongly interacting (large) dipoles.  The model templates incorporated contributions from the photon and Pomeron (elastic scattering) $p_T$ and for the detector resolution.

 The poor fit quality showed that this model cannot explain the data.  There are several possible apparent explanations, and it is likely that several of them contribute to the poor fit.  The small-dipole, Woods-Saxon model does not account for multiple interactions by a single dipole (i. e. as is accounted for by a Glauber calculation or in the dipole model); the Glauber calculation will alter the effective size of the nucleus.  The photon $p_T$ spectrum was also problematic, in that it was calculated for all impact parameters, rather than the actual limited range.  Earlier in the analysis chain, the dipole function used to fit and subtract the incoherent component likely fails at small $p_T$.      Many of these problems are also present in the Fourier-transform approach to finding the transverse gluon distributions.  The photon $p_T$ spectrum must be accurately known to be unfolded.  Multiple scattering changes the effective shape of the nucleus \cite{Frankfurt:2011cs}. 
 
Looking ahead, the LHC Run 3 should generate large samples of exclusive photoproduced $J/\psi$, without a trigger requirement for mutual Coulomb dissociation \cite{ Citron:2018lsq}.  This will reduce the photon $p_T$ spectrum uncertainties, and, more importantly, allow the rejection of most incoherent photoproduction via the rejection of events containing forward neutrons and protons.  This will greatly reduce the magnitude of the incoherent subtraction.
 
Most of these problems will be alleviated at the electron-ion collider \cite{AbdulKhalek:2021gbh}.   Except at small $Q^2$, the photon $p_T$ can be measured by observing the scattered electron, albeit with some uncertainty due to the imperfectly known electron initial momentum.  Critically, separation of coherent and incoherent production should be improved, since the detector far-forward subsystems will instrument almost all of phase space.

\section*{Acknowledgements}
Ya-Ping Xie made important contributions to the fitting effort. 

\paragraph{Funding information}
This work was funded by the U.S. DOE under contract number DE-AC02-05-CH11231.

\bibliography{KleinDIS2021.bib}

\begin{thebibliography}{10}
\providecommand{\url}[1]{\texttt{#1}}
\providecommand{\urlprefix}{URL }
\expandafter\ifx\csname urlstyle\endcsname\relax
  \providecommand{\doi}[1]{doi:\discretionary{}{}{}#1}\else
  \providecommand{\doi}{doi:\discretionary{}{}{}\begingroup
  \urlstyle{rm}\Url}\fi
\providecommand{\eprint}[2][]{\url{#2}}

\bibitem{Alvensleben:1970uw}
H.~Alvensleben \emph{et~al.},
\newblock \emph{{Photoproduction of neutral rho mesons from complex nuclei}},
\newblock Phys. Rev. Lett. \textbf{24}, 786 (1970),
\newblock \doi{10.1103/PhysRevLett.24.786}.

\bibitem{Frankfurt:2015cwa}
L.~Frankfurt, V.~Guzey, M.~Strikman and M.~Zhalov,
\newblock \emph{{Nuclear shadowing in photoproduction of \ensuremath{\rho}
  mesons in ultraperipheral nucleus collisions at RHIC and the LHC}},
\newblock Phys. Lett. B \textbf{752}, 51 (2016),
\newblock \doi{10.1016/j.physletb.2015.11.012},
\newblock \eprint{1506.07150}.

\bibitem{Acharya:2021ugn}
S.~Acharya \emph{et~al.},
\newblock \emph{{Coherent $\rm{J/\psi}$ and $\rm{\psi'}$ photoproduction at
  midrapidity in ultra-peripheral Pb-Pb collisions at
  $\sqrt{s_{\mathrm{NN}}}~=~5.02$ TeV}}  (2021),
\newblock \eprint{2101.04577}.

\bibitem{Good:1960ba}
M.~L. Good and W.~D. Walker,
\newblock \emph{{Diffraction disssociation of beam particles}},
\newblock Phys. Rev. \textbf{120}, 1857 (1960),
\newblock \doi{10.1103/PhysRev.120.1857}.

\bibitem{Klein:2019qfb}
S.~R. Klein and H.~M\"antysaari,
\newblock \emph{{Imaging the nucleus with high-energy photons}},
\newblock Nature Rev. Phys. \textbf{1}(11), 662 (2019),
\newblock \doi{10.1038/s42254-019-0107-6},
\newblock \eprint{1910.10858}.

\bibitem{Adamczyk:2017vfu}
L.~Adamczyk \emph{et~al.},
\newblock \emph{{Coherent diffractive photoproduction of $\rho^0$ mesons on
  gold nuclei at 200 GeV/nucleon-pair at the Relativistic Heavy Ion Collider}},
\newblock Phys. Rev. \textbf{C96}(5), 054904 (2017),
\newblock \doi{10.1103/PhysRevC.96.054904},
\newblock \eprint{1702.07705}.

\bibitem{Abelev:2007nb}
B.~I. Abelev \emph{et~al.},
\newblock \emph{{$\rho^0$ photoproduction in ultraperipheral relativistic heavy
  ion collisions at $\sqrt{s_{NN}}$ = 200 GeV}},
\newblock Phys. Rev. C \textbf{77}, 034910 (2008),
\newblock \doi{10.1103/PhysRevC.77.034910},
\newblock \eprint{0712.3320}.

\bibitem{Baltz:2002pp}
A.~J. Baltz, S.~R. Klein and J.~Nystrand,
\newblock \emph{{Coherent vector meson photoproduction with nuclear breakup in
  relativistic heavy ion collisions}},
\newblock Phys. Rev. Lett. \textbf{89}, 012301 (2002),
\newblock \doi{10.1103/PhysRevLett.89.012301},
\newblock \eprint{nucl-th/0205031}.

\bibitem{Baur:2003ar}
G.~Baur, K.~Hencken, A.~Aste, D.~Trautmann and S.~R. Klein,
\newblock \emph{{Multiphoton exchange processes in ultraperipheral relativistic
  heavy ion collisions}},
\newblock Nucl. Phys. A \textbf{729}, 787 (2003),
\newblock \doi{10.1016/j.nuclphysa.2003.09.006},
\newblock \eprint{nucl-th/0307031}.

\bibitem{Klein:2018grn}
S.~R. Klein,
\newblock \emph{{Dipion photoproduction and the $Q^2$ evolution of the shape of
  the gold nucleus}},
\newblock PoS \textbf{DIS2018}, 047 (2018),
\newblock \doi{10.22323/1.316.0047},
\newblock \eprint{1807.00455}.

\bibitem{Acharya:2021bnz}
S.~Acharya \emph{et~al.},
\newblock \emph{{First measurement of the |$t$|-dependence of coherent $J/\psi$
  photonuclear production}},
\newblock Phys. Lett. B \textbf{817}, 136280 (2021),
\newblock \doi{10.1016/j.physletb.2021.136280},
\newblock \eprint{2101.04623}.

\bibitem{Vidovic:1992ik}
M.~Vidovic, M.~Greiner, C.~Best and G.~Soff,
\newblock \emph{{Impact parameter dependence of the electromagnetic particle
  production in ultrarelativistic heavy ion collisions}},
\newblock Phys. Rev. C \textbf{47}, 2308 (1993),
\newblock \doi{10.1103/PhysRevC.47.2308}.

\bibitem{Klein:1999gv}
S.~R. Klein and J.~Nystrand,
\newblock \emph{{Interference in exclusive vector meson production in heavy ion
  collisions}},
\newblock Phys. Rev. Lett. \textbf{84}, 2330 (2000),
\newblock \doi{10.1103/PhysRevLett.84.2330},
\newblock \eprint{hep-ph/9909237}.

\bibitem{Klein:2020jom}
S.~Klein, A.~H. Mueller, B.-W. Xiao and F.~Yuan,
\newblock \emph{{Lepton Pair Production Through Two Photon Process in Heavy Ion
  Collisions}},
\newblock Phys. Rev. D \textbf{102}(9), 094013 (2020),
\newblock \doi{10.1103/PhysRevD.102.094013},
\newblock \eprint{2003.02947}.

\bibitem{Frankfurt:2011cs}
L.~Frankfurt, V.~Guzey and M.~Strikman,
\newblock \emph{{Leading Twist Nuclear Shadowing Phenomena in Hard Processes
  with Nuclei}},
\newblock Phys. Rept. \textbf{512}, 255 (2012),
\newblock \doi{10.1016/j.physrep.2011.12.002},
\newblock \eprint{1106.2091}.

\bibitem{Citron:2018lsq}
Z.~Citron \emph{et~al.},
\newblock \emph{{Report from Working Group 5}: {Future physics opportunities
  for high-density QCD at the LHC with heavy-ion and proton beams}},
\newblock CERN Yellow Rep. Monogr. \textbf{7}, 1159 (2019),
\newblock \doi{10.23731/CYRM-2019-007.1159},
\newblock \eprint{1812.06772}.

\bibitem{AbdulKhalek:2021gbh}
R.~Abdul~Khalek \emph{et~al.},
\newblock \emph{{Science Requirements and Detector Concepts for the
  Electron-Ion Collider: EIC Yellow Report}}  (2021),
\newblock \eprint{2103.05419}.

\end{thebibliography}

\nolinenumbers

\end{document}